\documentclass{article} 
\usepackage{iclr2020_conference,times}

\usepackage{amsmath,amsfonts,bm}

\def\eqref#1{equation~\ref{#1}}

\def\1{\bm{1}}

\DeclareMathAlphabet{\mathsfit}{\encodingdefault}{\sfdefault}{m}{sl}
\SetMathAlphabet{\mathsfit}{bold}{\encodingdefault}{\sfdefault}{bx}{n}

\usepackage{hyperref}
\usepackage{url}
\usepackage{enumitem}

\makeatletter
\renewenvironment{description}%
               {\list{}{\leftmargin=18pt 
                        \labelwidth\z@ \itemindent-\leftmargin
                        }}%
               {\endlist}
\makeatother

\title{Levels of Analysis for Machine Learning}

\author{Jessica B.~Hamrick\\
DeepMind \\
\texttt{jhamrick@google.com} \\
\And
Shakir Mohamed\\
DeepMind \\
\texttt{shakir@google.com} \\
}

\iclrfinalcopy 
\begin{document}

\maketitle

\begin{abstract}
Machine learning is currently involved in some of the most vigorous debates it has ever seen.
Such debates often seem to go around in circles, reaching no conclusion or resolution.
This is perhaps unsurprising given that researchers in machine learning come to these discussions with very different frames of reference, making it challenging for them to align perspectives and find common ground.
As a remedy for this dilemma, we advocate for the adoption of a common conceptual framework which can be used to understand, analyze, and discuss research.
We present one such framework which is popular in cognitive science and neuroscience and which we believe has great utility in machine learning as well: \emph{Marr's levels of analysis}.
Through a series of case studies, we demonstrate how the levels facilitate an understanding and dissection of several methods from machine learning.
By adopting the levels of analysis in one's own work, we argue that researchers can be better equipped to engage in the debates necessary to drive forward progress in our field.
\end{abstract}

\vspace{-2mm}
\section{Conceptual Frameworks}

The last few months and years have seen researchers embroiled in heated debates touching on fundamental questions in machine learning: what, exactly, is ``deep learning''?
Can it ever be considered ``symbolic''?
Is it sufficient to achieve artificial general intelligence?
Can it be trusted in high-stakes, real-world applications? 
These are important questions, yet the discussions surrounding them seem unable to move forward productively.
This is due, in part, to the lack of a common framework within which argument and evidence are grounded and compared.
Early researchers in computer science, engineering and cognitive science struggled in this same way, and were led to  develop several \emph{conceptual frameworks}: mental models that align researchers' understanding of how their work and philosophies fit into the larger goals of their field and science more broadly \citep[e.g.][]{chomsky1965aspects,marr1976understanding,newell1976computer,marr1982vision,newell1981knowledge,arbib1987levels,anderson1990adaptive}.
Our proposition is to reinvigorate this former tradition of debate by making the use of conceptual frameworks a core part of machine learning research.

We focus in this paper on the use of \emph{Marr's levels of analysis}, a conceptual framework popular in cognitive science and neuroscience.
\citet{marr1982vision} identified a three-layer hierarchy for describing and analyzing a computational system:
\vspace{-2mm}
\begin{description}
\setlength{\itemsep}{1pt}
  \setlength{\parskip}{0pt}
  \setlength{\parsep}{0pt}
\item[Computational level.] \emph{What is the goal of a system, what are its inputs and outputs, and what mathematical language can be used to express that goal?} As an example, let us consider the case of natural language. In one view, \citet{chomsky1965aspects} famously argued that the purpose of language is to structure thought by expressing knowledge in a compositional and recursive way.
\item[Algorithmic or representational level.] \emph{What representations and algorithms are used to achieve the computational-level goal?} Under Chomsky's view of language, we might hypothesize that a language processing system uses symbolic representations like syntax trees and that those representations are formed using some form of linguistic parsing.
\item[Implementation level.] \emph{How is the system implemented, either physically or in software?} For example, we might ask how a neural circuit in the brain could represent syntax trees using population-based codes, or how a parser should be implemented efficiently in code.
\end{description}

Each level of analysis allows us to identify various hypotheses and constraints about how the system operates, and to highlight areas of disagreement.
For example, others have argued that the computational-level goal of language is not to structure thought, but to communicate with others \citep{tomasello2010origins}.
At the algorithmic level, we might argue that syntax trees are an impoverished representation for language, and that we must also consider dependency structures, information compression schemes, context, pragmatics, and so on.
And, at the implementation level, we might consider how things like working memory or attention constrain what algorithms can be realistically supported.

\section{Case Studies in Using Levels of Analysis}

We use four case studies to show how Marr's levels of analysis can be applied to understanding methods in machine learning, choosing examples that span the breadth of research in the field.
While the levels have traditionally been used to reverse-engineer biological behavior, here we use them as a conceptual device that facilitates the structuring and comparison of machine learning methods.
Additionally, we note that our classification of various methods at different levels of analysis may not be the only possibility, and readers may find themselves disagreeing with us.
This disagreement is a desired outcome: the act of applying the levels forces differing assumptions out into the open so they can be more readily discussed.

\vspace{-2mm}
\subsection{Deep Q-Networks (DQN)}

DQN \citep{mnih2015human} is a deep reinforcement learning algorithm originally used to train agents to play Atari games.
At the \textbf{computational} level, the goal of DQN is to efficiently produce actions which maximize scalar rewards given observations from the environment.
We can use the language of reinforcement learning---and in particular, the Bellman equation---to express this goal \citep{sutton2018reinforcement}.
At the \textbf{algorithmic} level, DQN performs off-policy one-step temporal difference (TD) learning (i.e., Q-learning), which can be contrasted with other choices such as SARSA, $n$-step Q-learning, TD($\lambda$), and so on.
All of these different algorithmic choices are different realizations of the same goal: to solve the Bellman equation.
At the \textbf{implementation} level, we consider how Q-learning should actually be implemented in software.
For example, we can consider the neural network architecture, optimizer, and hyperparameters, as well as components such as a target network and replay buffer.
We could also consider how to implement Q-learning in a distributed manner \citep{horgan2018distributed}.

The levels of analysis when applied to machine learning systems help us to more readily identify assumptions that are being made and to critique those assumptions.
For example, perhaps we wish to critique DQN for failing to \emph{really} know what objects are, such as what a ``paddle'' or ``ball'' is in Breakout \citep{marcus2018deep}.
While on the surface this comes across as a critique of deep reinforcement learning more generally, by examining the levels of analysis, it becomes clear that this critique could actually be made in very different ways at two levels of analysis: \vspace{-2mm}
\begin{enumerate}[leftmargin=*]
\setlength{\itemsep}{1pt}
  \setlength{\parskip}{0pt}
  \setlength{\parsep}{0pt}
\item At the computational level, the goal of DQN is not to learn about objects but to maximize scalar reward in a game.
Yet, if we care about the system understanding objects, then perhaps the goal should be formulated such that discovering objects is part of the objective itself \citep[e.g.][]{burgess2019monet,greff2019multi}.
Importantly, the fact that the agent uses deep learning is orthogonal to the question of what the goal is.
\item At the implementation level, we might interpret the critique as being about the method used to implement the Q-function.
Here, it becomes relevant to talk about deep learning.
DQN uses a convolutional neural network to convert visual observations into distributed representations, but we could argue that these are not the most appropriate for capturing the discrete and compositional nature of objects  \citep[e.g.][]{battaglia2018relational,van2019perspective}.
\end{enumerate}
Depending on whether the critique is interpreted at the computational level or implementation level, one might receive very different responses, thus leading to a breakdown of communication.
However, if the levels were to be used and explicitly referred to when making an argument, there would be far less room for confusion or misinterpretation and far more room for productive discourse.

\vspace{-2mm}
\subsection{Convolutional neural networks}

One of the benefits of the levels of analysis is that they can be flexibly applied to any computational system, even when that particular system might itself be a component of a larger system.
For example, an important component of the DQN agent at the implementation level is a visual frontend which processes observations and maps them to Q-values \citep{mnih2015human}.
Just as we can analyze the whole DQN agent using the levels of analysis, we can do the same for DQN's visual system.

At the \textbf{computational} level, the goal of a vision module is to map spatial data (as opposed to, say, sequential data or graph-structured data) to a compact representation of objects and scenes that is invariant to certain types of transformations (such as translation).
At the \textbf{algorithmic} level, feed-forward convolutional networks \citep{fukushima1988neocognitron,lecun1989backpropagation} are one type of procedure that processes spatial data while maintaining translational invariance.
Within the class of CNNs, there are many different versions, such as AlexNet \citep{krizhevsky2012imagenet}, ResNet \citep{he2016deep}, or dilated convolutions \citep{yu2015multi}.
However, this need not be the only choice.
For example, we could consider a relational \citep{wang2018non} or autoregressive \citep{oord2016pixel} architecture instead.
Finally, at the \textbf{implementation} level, we can ask how to efficiently implement the convolution operation in hardware.
We could choose to implement our CNN on a CPU, a single GPU, or multiple GPUs.
We may also be concerned with questions about what to do if the parameters or gradients of our network are too large to fit into GPU memory, how to scale to much larger batch sizes, and how to reduce latency.

Analyzing the CNN on its own highlights how the levels can be applied flexibly to many types of methods: we can ``zoom'' our analysis in and out to focus understanding and discussion of different aspects of larger systems as needed.
This ability is particularly useful for analyzing whether a component of a larger system is really the right one by comparing the role of a component in a larger system to its computational-level goal in isolation.
For example, as we concluded above, the computational-level goal of a CNN is to process spatial data while maintaining translational invariance.
This might be appropriate for certain types of goals (e.g., object classification) but not for others in which translational invariance is inappropriate (e.g., object localization).

\subsection{Symbolic Reasoning on Graphs}

A major topic of debate has been the relationship between symbolic reasoning systems and distributed (deep) learning systems.
The levels of analysis provide an ideal way to better illuminate the form of this relationship.
As an illustrative example, let us consider the problem of solving an NP-complete problem like the Traveling Salesman Problem (TSP), a task that has traditionally been approached symbolically.
At the \textbf{computational} level, given a complete weighted graph (i.e. fully-connected edges with weights), the goal is to find the minimum-weight path through the graph that visits each node exactly once.
Although finding an exact solution to the TSP takes (in the worst case) exponential time, we could formulate part of the goal as finding an efficient solution which returns near-optimal or approximate solutions.
At the \textbf{algorithmic} level, there are many ways we could try to both represent and solve the TSP.
As it is an NP-complete problem, we could choose to transform it into other types of NP-complete problems (such as graph coloring, the knapsack problem, or boolean satisfiablity).
A variety of algorithms exist for solving these different types of problems, with a common approach being heuristic search.
While heuristics are typically handcrafted at the \textbf{implementation} level using a symbolic programming language, they could also be implemented using deep learning components \citep[e.g.][]{vinyals2015pointer,dai2017learning}.

Performing this analysis shows how machine learning systems may simultaneously implement computations with \emph{both} symbolic and distributed representations at different levels of analysis.
Perhaps, then, discussions about ``symbolic reasoning'' versus ``deep learning'' or (``hybrid systems'' versus ``non-hybrid systems'') are not the right focus because \emph{both} symbolic reasoning and deep learning already coexist in many architectures.
Instead, we can use the levels to productively steer discussions to where they can have more of an impact.
For example, we could ask: is the logic of the algorithmic level the right one to achieve the computational-level goal?
Is that goal the one we care about solving in the first place?
And, is the architecture used at the implementation level appropriate for learning or implementing the desired algorithm?

\vspace{-2mm}
\subsection{Machine Learning in Healthcare}

The last few case studies have focused on applying Marr's levels to widely-known algorithms.
However, an important component of any machine learning system is also the data on which it is trained and its domain of application.
We turn to an application of machine learning in real-world settings where data plays a significant  role.
Consider a clinical support tool for managing electronic health records (EHR) \citep[e.g.][]{wu2010prediction, tomavsev2019clinically}.
At the \textbf{computational} level, the goal of this system is to improve patient outcomes by predicting patient deterioration during the course of hospitalization. 
Performing this computational-level analysis encourages us to think deeply about all facets of the overall goal, such as: what counts as a meaningful prediction that gives sufficient time to act?
Are there clinically-motivated labels available?
How should the system deal with imbalances in clinical treatments and patient populations?
What unintended or undesirable consequences might arise from deployment of such a system?
The \textbf{algorithmic} level asks what procedures should be used for collecting and representing the data in order to achieve the computational-level goal.
For example, how are data values encoded (e.g., censoring extreme values, one-hot encodings, using derived summary-statistics)?
Is the time of an event represented using regular intervals or not?
And, what methods allow us to develop longitudinal models of such high-dimensional and highly-sparse data?
This is where familiar models of gradient-boosted trees or logistic regression enter.
Finally, at the \textbf{implementation} level, questions arise regarding how the system is deployed.
For example, we may need to consider data sandboxing and security, inter-operability, the role of open-standards, data encryption, and interactions and behavioural changes when the system is deployed.

In addition to facilitating debate, applying Marr's levels of analysis in a domain like healthcare emphasizes important aspects of design decisions that need to be made.
For example, the representation we choose at the algorithmic level can lock in choices at the beginning of a research program that has important implications for lower levels of analysis. 
Choosing to align the times when data is collected into six hour buckets will affect underlying structure and predictions, with implications on how performance is analyzed and for later deployment.
Similarly, ``alerting fatigue'' is a common concern of such methods, and can be debated at the computational level by asking questions related to the broader interactions of the system with existing clinical pathways, or at the implementation level through mechanisms for alert-filtering.
When we consider questions of bias in the data sets used, especially in healthcare, the levels can help us identify whether such bias is a result of clinical constraints at the computational level, or whether it is an artefact of algorithmic-level choices.

\vspace{-2mm}
\section{Beyond Marr}

As our case studies have demonstrated, Marr's levels of analysis are a powerful conceptual framework that can be used to define, understand, and discuss research in machine learning.
The reason that the levels work so well is that they allow researchers to  approach discussion using a common frame-of-reference that highlights different choices or assumptions, streamlining discussion while also encouraging skepticism.
We suggest readers try this exercise themselves: apply the levels to a recent paper that you enjoyed (or disliked!) as well as your own research.
We have found this approach to be a useful way to understand our own specific papers as well as the wider research within which they sit.
While the process is sometimes harder than expected, the difficulties that arise are are often exactly in the places we have overlooked and where deeper investigation is needed.

Machine learning research often implicitly relies on several similar conceptual frameworks, although they are not often formalized in the way Marr's levels are. 
For example, most deep learning researchers will recognize an \textit{architecture-loss} paradigm: a given problem is studied in terms of a computational graph that processes and transforms data into some output, and a method that computes prediction errors and propagates them to drive parameter and distribution updates.
Similarly, many software engineers will recognize the \textit{algorithm-software-hardware} distinction: a problem can be understood in terms of the algorithm to be computed, the software which computes that algorithm, and the hardware on which the software runs.
While both of these frameworks are conceptually useful, neither fully captures the full set of abstractions ranging from the computational-level goal to the physical implementation.

Of course, Marr's levels are not without limitations or alternatives.
Even in the course of writing this paper, we considered whether it would make sense to split the algorithmic level into two, mirroring the original four-level formulation put forth by \citet{marr1976understanding}.
Researchers from cognitive science and neuroscience have also spent considerable effort discussing alternate formulations of the levels \citep[e.g.][]{sun2008introduction,griffiths2012bridging,poggio2012levels,danks2013moving,peebles2015thirty,niv2016reinforcement,yamins2016eight,krafft2018levels,love2019levels}.
For example, \citet{poggio2012levels} argues that \textit{learning} should be added as an additional level within Marr's hierarchy.
Additionally, Marr's levels do not explicitly recognize the socially-situated role of computational systems, leading to proposals for alternate hierarchies that consider the interactions between computational systems themselves, as well as the socio-cultural processes in which those systems are embedded \citep{nissenbaum2001computer,sun2008introduction}.
The ideal conceptual framework for machine learning will need to go beyond Marr to also address such considerations.
In the meantime, we believe that Marr's levels are useful as a common conceptual lens for machine learning researchers to start with.
Given such a lens, we believe that researchers will be better equipped to discuss and view their research, and to ultimately address the many deep questions of our field.

\section*{Acknowledgments}

We would like to thank M. Overlan, K. Stachenfeld, T. Pfaff, A. Santoro, D. Pfau, A. Sanchez-Gonzalez, M. Rosca, and two anonymous reviewers for helpful comments and feedback on this paper. Additionally, we would like to thank I. Momennejad, A. Erdem, A. Lampinen, F. Behbahani, R. Patel, P. Krafft, H. Ritz, T. Darmetko, and many others on Twitter for providing a variety of interesting perspectives and thoughts on this topic that helped to inspire this paper.

\bibliography{iclr2020_conference}
\bibliographystyle{iclr2020_conference}

\end{document}